\documentclass[aps,prc,twocolumn,superscriptaddress]{revtex4}
\usepackage{graphicx}
\usepackage{dcolumn}
\usepackage{bm}
\usepackage{siunitx}

\begin{document}
\title{Precise measurement of the thermal and stellar $^{54}$Fe($n, \gamma$)$^{55}$Fe cross sections via AMS}

\author{A. Wallner} \email[corresponding author: ]{anton.wallner@anu.edu.au} 
 \affiliation{Department of Nuclear Physics, Research School of Physics and Engineering, 
The Australian National University, Canberra, ACT 2601, Australia}, 
\affiliation{VERA Laboratory, Faculty of Physics, University of Vienna, Austria}
\author{T. Belgya} \affiliation{Nuclear Analysis and Radiography Department, Institute for Energy Security and Environmental Safety, Centre for Energy Research, Hungarian Academy of Sciences}
\author{K. Buczak} 
\affiliation{VERA Laboratory, Faculty of Physics, University of Vienna, Austria}
\author{L. Coquard} \affiliation{Karlsruhe Institute of Technology (KIT), Campus North, Institute of Nuclear 
	Physics, PO Box 3640, Karlsruhe, Germany}
\author{M. Bichler} \affiliation{Atominstitut, Vienna University of Technology, Austria}
\author{I. Dillmann} \altaffiliation[Present address: ]{TRIUMF, Vancouver BC, Canada}
	\affiliation{Karlsruhe Institute of Technology (KIT), Campus North, Institute of Nuclear 
	Physics, PO Box 3640, Karlsruhe, Germany}
\author{R. Golser} 
\affiliation{VERA Laboratory, Faculty of Physics, University of Vienna, Austria}
\author{F.~K{\"a}ppeler} \affiliation{Karlsruhe Institute of Technology (KIT), Campus North, 
	Institute of Nuclear Physics, PO Box 3640, Karlsruhe, Germany}
\author{A.~Karakas} \affiliation{Monash Centre for Astrophysics, School of Physics and Astronomy, Monash University, VIC 3800, Australia},
\affiliation{Research School of Astronomy and Astrophysics, The Australian National University, Canberra,  ACT 2611, Australia}
\author{W. Kutschera} \affiliation{VERA Laboratory, Faculty of Physics, University of Vienna, Austria}
\author{C. Lederer}  \affiliation{School of Physics and Astronomy, University of Edinburgh, UK}
	{\affiliation{VERA Laboratory, Faculty of Physics, University of Vienna, Austria}
\author{A. Mengoni} \affiliation{CERN, CH-1211 Geneva 23, Switzerland}
\author{M.~Pignatari} \altaffiliation{NuGrid collaboration, \url{http://www.nugridstars.org}}
\affiliation{ E.A.~Milne Centre for Astrophysics, Dept. of Physics \& Mathematics, University of Hull, United Kingdom}  
\author{A. Priller} \affiliation{VERA Laboratory, Faculty of Physics, University of Vienna, Austria} 
\author{R. Reifarth} \affiliation{Institute of Applied Physics, Goethe University Frankfurt, Frankfurt, Germany}
\author{P. Steier} \affiliation{VERA Laboratory, Faculty of Physics, University of Vienna, Austria}
\author{L. Szentmiklosi} \affiliation{Nuclear Analysis and Radiography Department, Institute for Energy Security and Environmental Safety, Centre for Energy Research, Hungarian Academy of Sciences}
\date{\today}

\begin{abstract}
The detection of long-lived radionuclides through ultra-sensitive single atom counting
via accelerator mass spectrometry (AMS) offers opportunities for precise measurements of
neutron capture cross sections, e.g. for nuclear astrophysics. The technique represents a truly 
complementary approach, completely independent of previous experimental methods. The 
potential of this technique is highlighted at the example of the  $^{54}$Fe($n, \gamma$)$^{55}$Fe 
reaction. Following a series of irradiations with neutrons from cold and thermal to keV energies, the 
produced long-lived $^{55}$Fe nuclei ($t_{1/2}=2.744(9)$ yr) were analyzed at the Vienna 
Environmental Research Accelerator (VERA). A reproducibility of about 1\% could be achieved 
for the detection of $^{55}$Fe, yielding cross section uncertainties of less than 3\%. Thus, the new data can serve as anchor points to time-of-flight experiments. We report
significantly improved neutron capture cross sections at thermal energy ($\sigma_{th}=2.30\pm0.07$ 
b) as well as for a quasi-Maxwellian spectrum of $kT=25$ keV ($\sigma=30.3\pm1.2$ mb) and 
for $E_n=481\pm53$ keV ($\sigma= 6.01\pm0.23$ mb). The new experimental cross sections have been 
used to deduce improved Maxwellian average cross sections in the temperature regime of
the common $s$-process scenarios. The astrophysical impact is discussed using stellar models 
for low-mass AGB stars. 
\end{abstract}
\maketitle

\section{Introduction \label{sec:1}}

An increasing number of abundance observations in very rare, ultra metal-poor (UMP) stars 
in the galactic halo indicates abundance patterns that scale approximately with the solar $r$
component for elements heavier than barium \cite{SCG08}, but with star-to-star variations 
questioning the paradigm of a robust $r$-process production \cite{RST10}. For lighter 
elements, there are significant discrepancies. Differences of the order of 20\% are also found 
between the solar $s$-process abundances in the mass range $90\leq A \leq 140$ and the 
results of Galactic chemical evolution studies \cite{TGA04}. This result for the $s$ process is 
mainly due to the achievements of nuclear astrophysics in the past decades \cite{RLK14}. 
However, as large stellar physics uncertainties are still affecting theoretical predictions of the 
$s$ process, a set of precise experimental nuclear reaction rates is a fundamental requirement 
to tackle these challenges. Further improvements in the standard prescriptions of $s$- and/or 
$r$-process nucleosynthesis are clearly needed for a refined view on the origin and enrichment 
of the elements in the Universe.

In the course of these investigations the $s$ process plays a key role because the $s$ 
abundances can be reliably quantified and in turn serve to derive the $r$ abundances via the
residual method \cite{KGB11}. To fully exploit the potential of the $s$ process as an 
abundance reference, it is necessary to establish an accurate set
of the underlying nuclear physics data. In this context, neutron capture cross sections in 
the keV energy range are particularly important because of their strict correlation with the 
emerging $s$ abundances and their effect on the overall neutron balance. 

Most of the $^{54}$Fe in the universe is made by explosive Si- and O-burning in core-collapse 
supernovae \cite{WHW02} and in thermonuclear supernovae \cite{HiN00}. $^{54}$Fe is not 
produced in the $s$ process, but is instead depleted by neutron capture, according to its cross 
section. Small amounts of $^{54}$Fe can be potentially produced by different types of $p$ 
processes (e.g., \cite{RDD13, PZH15}), but with negligible relevance for the galactic inventory.

Indications of neutron capture on $^{54}$Fe have been found via isotopic ratios in different 
types of presolar SiC grains that condensed in supernovae ejecta and in the envelopes of low 
mass AGB stars and were trapped in pristine meteorites in the early solar system \cite{MAG08}. 
In these grains, Fe isotopic abundances are composed of normal pristine material and stellar 
matter processed by neutron capture. While the normal material carries the signature of galactic 
chemical evolution, the stellar material is determined by the respective ($n, \gamma$) cross 
sections, which are, therefore, crucial for quantitative analyses.  
 
The information on keV-neutron capture cross sections has been summarized in
compilations of Maxwellian-averaged cross sections (MACS) for $s$-process applications 
\cite{BBK00,DPK09,DPK14}. In spite of the numerous data in literature, these collections 
clearly exhibit the need for significant MACS improvements to resolve discrepancies and/or to
reach the necessary accuracy of 2-5\% \cite{KGB11} by dedicated precision measurements. 

The present study of the $^{54}$Fe($n, \gamma$)$^{55}$Fe cross section is motivated by 
these aspects, i.e. to remove previous discrepancies and to provide a sensitive test for the
 treatment of broad s-wave resonances in the analysis of time-of-flight (TOF) 
experiments. The proper strength of such resonances, which can dominate the MACS values 
in typical $s$-process environments, are difficult to extract from measured data. Because
of their very large neutron widths, the scattering probability exceeds the capture channel by
orders of magnitude, and the corrections for the effect of scattered neutrons are often 
obscuring the capture signal \cite{KWG00}. This inherent problem of TOF measurements, 
which has to be treated by extensive simulations of the particular experimental situation 
\cite{ZBC16}, is avoided in careful activation measurements. 

The case of $^{54}$Fe is appealing because the activation method can be combined with 
accelerator mass spectrometry (AMS) for detecting directly the $^{55}$Fe nuclei produced 
in the capture reaction. This technique provides a powerful complement of the activation 
method as it is essentially independent of the half-life and decay characteristics of the reaction 
product, thus reducing the related uncertainties of the traditional activity technique
\cite{NPA05a, Wal10, WBB14}}. Another advantage is that AMS requires only small sample masses of order of mg, thus scattering corrections inherent to TOF measurements are completely avoided.

The paper is organized in the following way: Existing data in the literature are 
summarized in Sec. \ref{sec:2}. The following Secs. \ref{sec:3} and 
\ref{sec:4} are dealing with the neutron activations and the AMS 
measurements. Data analysis and results are presented in Sec. \ref{sec:5},
the astrophysical aspects are discussed in Sec. \ref{sec:astro}, and a
summary is given in Sec. \ref{sec:sum}. 

\section{Previous data\label{sec:2}}

The present experiment is the first attempt to determine the $^{54}$Fe($n, \gamma$) 
cross section at keV energies via the activation method. This method had not been used so 
far because the very weak signals from the EC decay of $^{55}$Fe are difficult
to detect quantitatively. All previous data were, therefore, obtained by TOF 
measurements, starting with the work of Beer and Spencer \cite{BeS75}, who 
reported capture and transmission data in the energy range 5 to 200 keV and 10 
to 300 keV, respectively, but were missing the important s-wave resonance at 
7.76 keV, which contributes about 30\% to the MACS value at $kT=30$ keV.
Therefore, these results have been omitted in the further discussion.

The first complete list of capture kernels $k_\gamma=g\Gamma_n\Gamma_\gamma/\Gamma$ 
in the astrophysically relevant energy from 0.1 to 500 keV was obtained by Allen 
{\it et al.} at OLELA (Oakridge, ORNL) \cite{AMB77,AMB79}. As this measurement was carried out with a rather 
thick sample of 2 at/barn, neutron multiple scattering and the detector response to 
scattered neutrons were causing significant background effects. For the broad 
s-wave resonances below 100 keV, which dominate the stellar cross section of 
$^{54}$Fe, large corrections of up to 30 and 50\% had to be considered for
these effects, respectively.

These corrections could be considerably reduced in a subsequent measurement 
by Brusegan {\it et al.} at GELINA (JRC/IRMM, Geel) \cite{BCR83}. With a much thinner sample of only 0.023 
at/barn, the set of capture kernels could be significantly improved in the investigated
neutron energy range below 200 keV.

Recently, Giubrone and the n\_TOF collaboration \cite{Giu14,GDT14} took advantage 
of the intense, high-resolution neutron source at CERN for further improving the
capture data of $^{54}$Fe from thermal to 500 keV. By reducing the sample dimensions 
again by factors of 3 and 25 in thickness and mass, respectively, and by application of
refined analysis methods the set of resonance parameters could be obtained 
with unprecedented accuracy. 

For a thermal energy of $kT=30$ keV the MACS values deduced from these 
TOF measurements are compared in Table \ref{tab:comp} with data from the previous
KADoNiS v0.3 compilation (www.kadonis.org) as well as with the results calculated from the 
evaluated cross sections in the main data libraries ENDF/B-VII.1 \cite{CHO11},
JENDL-4.0 \cite{SIN11}, and JEFF-3.2 \cite{JEF14}. The KADoNiS 
value represents an average of the older TOF measurements \cite{BCR83} and 
\cite{AMB77,AMB79}. In view of the consistent results of the refined measurements
\cite{Giu14, BCR83} it is surprising to find that the MACS values obtained with the 
evaluated cross sections are about 30\% smaller. This situation clearly underlines 
the need for the present measurement, which is based on a completely 
independent experimental technique.

\begin{table}[htb]
\caption{Comparison of previous experimental results for the MACS of 
$^{54}$Fe at $kT=30$ with data from compilation and major data libraries.
\label{tab:comp}}
\begin{ruledtabular}
\begin{tabular} {lc}
 	Data from 							& $\langle \sigma v \rangle$/$v_T$ (mb)	\\
\hline
n\_TOF (exp)		\cite{Giu14,GDT14}	& $28.5\pm1.6$			\\
GELINA (exp)		\cite{BCR83}			& $27.6\pm1.8$			\\
ORNL (exp)			\cite{AMB77,AMB79}	& $33.6\pm2.7$			\\
										&						\\
KADoNiS (comp) 	\cite{DPK09}			& $29.6\pm1.3$			\\ 
ENDF-B/VII.1 (eval)	\cite{CHO11}		& $21.6\pm2.7$			\\ 
JENDL-4.0	 (eval)	\cite{SIN11}			& $21.6$				\\ 
JEFF-3.2 (eval)		\cite{JEF14}			& $21.6$				\\ 
\end{tabular}
\end{ruledtabular}
\end{table}

With respect to the thermal cross section value, according to the compilation of Mughabghab \cite{Mug06} the thermal cross 
section of $\sigma_{th}=2.25\pm0.18$ b exhibits a comparably large 
uncertainty of 8\% which again reflects the difficulty of quantifying the reaction product  $^{55}$Fe. We developed our measurement technique first with activations at cold and thermal neutron energies as they exhibit about 100 times higher neutron capture cross sections compared to keV energies.

\section{Neutron irradiations \label{sec:3}}

\subsection{Activations with thermal neutrons at ATI Vienna \label{sec:3.1}}

The activations with thermal neutrons ($kT=25$ meV; 300 K) were performed 
at the TRIGA Mark-II reactor at the Atominstitut in Vienna (ATI) in a 
well-characterized thermal spectrum. The neutron flux at the irradiation position 
about 1 m from the core was $\sim3.7\times10^{11}$ cm$^{-2}$s$^{-1}$.
This position provides a thermal to epithermal flux ratio of ~76 (verified via the 
Zr standard method \cite{VBW08}). 

In total, four irradiations between 1 and 10 minutes were performed using Zr 
foils as flux monitors (Table \ref{tab:irrth}. The Fe samples were prepared from metal powder of 
natural isotopic composition. The isotope 
composition of the natural material ($5.845\pm0.035$\% $^{54}$Fe, 
$91.754\pm0.036$ $^{56}$Fe) was adopted from Ref. \cite{BeW11}.
An amount of about 500 mg Fe powder, which was 
acquired from two different providers (Merck and Alfa Aesar), was encapsulated 
in plastic vials. The neutron fluence was determined by means of Zr foils 
attached to the vials via the induced $^{95}$Zr activity, using the thermal cross section value for $^{94}$Zr($n, \gamma$) of ($0.0494\pm0.0017$) barn \cite{Mug06}. The activities of these foils indicated flux variations of up 
to 5\%  between different activations.  

The epithermal contribution to the  $^{54}$Fe($n, \gamma$) cross section was 
only about 1\%, but the 7\% correction required for determining the flux with the $^{94}$Zr($n, \gamma$) 
reference cross section had a significant effect on the uncertainty of the ATI 
result and was limiting the accuracy of the ATI fluence to about 5\% (Sec.
\ref{sec:5.1}).

\begin{table}[htb]
\caption{Irradiations at thermal neutron energies.
\label{tab:irrth}}
\begin{ruledtabular}
\begin{tabular} {lccc}
Sample		& 	Irradiation  	& Neutron flux 				&	Monitor: thermal 	 		\\
			&	time (s)	&($10^{11}$cm$^{-2}$s$^{-1}$) 			& cross section (mb)					\\
\hline
\\[-0.7em]
ATI-FeM		& 600			& $3.53\pm0.18$		& $^{94}$Zr($n, \gamma$): $49.4\pm1.7$	\\
ATI-Fe2		& 600			& $3.73\pm0.20$		& 	\\
ATI-FeA2	 & 120			& $3.83\pm0.29$		&	\\
ATI-FeA4	 &  73		      & $3.62\pm0.19$		& \\
\end{tabular}
\end{ruledtabular}
\end{table}
				
\subsection{Activations with cold neutrons at the BNC Budapest  \label{sec:3.2}}

The irradiations were conducted at the 10-MW research reactor of the Budapest 
Neutron Centre (BNC) using the facilities for prompt gamma activation analysis (PGAA)
and the neutron-induced prompt gamma-ray spectrometer (NIPS) \cite{RBK04, SBR10, BKS14}. 
The neutrons from the reactor core were transported in a neutron guide tube 
resulting in a cold neutron beam with an average neutron energy of 10 meV. The typical neutron 
flux at the irradiation position was 3 and 4$\times10^7$ cm$^{-2}$s$^{-1}$ 
(thermal equivalent) for the NIPS and PGAA station, respectively.

Two iron samples 6 mm in diameter were prepared: one consisting of 44 mg
metal powder of natural isotopic composition and the second of almost pure 
$^{54}$Fe (45.2 mg, 99.85\% enrichment, STB Isotope GmbH). Approximately 
20 mg of Au powder were homogeneously mixed with the iron powder and the 
mixture was pressed into pellets. The pellets were then sandwiched by two Au foils 
of the same diameter forming a stack Au-(Fe/Au)-Au. 

The Au foils and the Au powder in the iron matrix were used to deduce the thermal 
equivalent neutron fluence in the irradiations, which lasted for about 1 and 4 d, 
respectively (Table \ref{tab:irrcold}) \cite{B09}. The fluence was 
determined from the induced $^{198}$Au activity of the monitor foils using the thermal cross section value for $^{197}$Au($n, \gamma$) of $98.65\pm0.09$ barn. 

\begin{table}[htb]
\caption{Irradiations at cold neutron energies.
\label{tab:irrcold}}
\begin{ruledtabular}
\begin{tabular} {lccc}
Sample		& 	Irradiation  	& Neutron flux 		 	&	Monitor: thermal 			\\
			&	time (min)	&($10^{7}$cm$^{-2}$s$^{-1}$) 		& 	cross section (b)				\\
\hline	
\\[-0.7em]	
BNC-FeM		& 5449		& $3.19\pm0.07$	& $^{197}$Au($n, \gamma$):$98.65\pm0.09$ 	\\
BNC-Fe4		& 1481		& $4.46\pm0.10$	&  	\\
\end{tabular}
\end{ruledtabular}
\end{table}

\subsection{Activations with keV neutrons \label{sec:3.3}}

\begin{figure}[ht]
  \includegraphics[width=8cm]{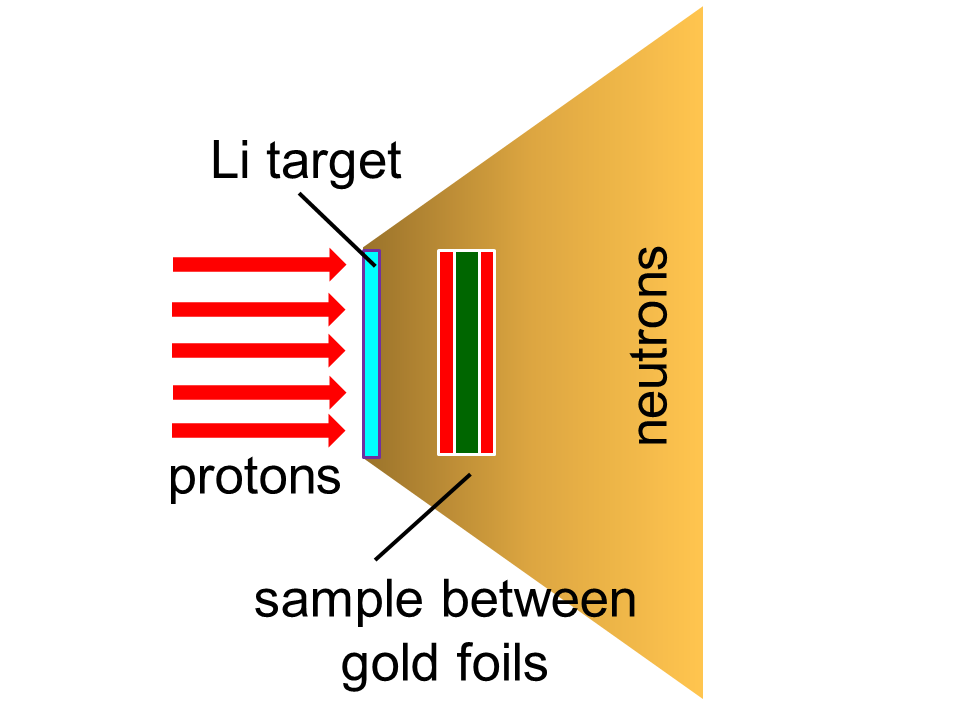}
 \caption{Color online. Schematic sketch of the setup used for the neutron
activations at the Karlsuhe Van de Graaff. \label{fig:setup}}
\end{figure}

The irradiations with keV neutrons were carried out at the Karlsruhe Institute 
of Technology (KIT) using the 3.7~MV Van de Graaff accelerator. Neutrons were 
produced via the $^7$Li($p, n$)$^7$Be reaction by bombarding 5 and 
30~$\mu$m thick layers of metallic Li on a 1 mm thick water-cooled 
Cu backing with proton beam currents of 80-90 $\mu$A. The thickness of the 
Li layers was controlled by means of a calibrated oszillating quartz monitor. 
During the irradiations, the neutron flux history was registered in intervals 
of 90~s by a $^6$Li-glass detector in 1~m distance from the neutron target. 
With this information it is possible to correct the fraction of decays during 
irradiations properly, including the fact that the Li targets degrade during 
the activation. A schematic sketch of the experimental setup is shown in 
Fig. \ref{fig:setup}.

Two sets of Fe samples from two different providers (see Sec. \ref{sec:3.1}) were prepared by pressing high-purity metal powder of 
natural isotopic composition into thin pellets 6 mm in diameter. 
During the activations the Fe samples were sandwiched between thin gold foils of the 
same diameter. The sample properties are summarized in Table \ref{samples}.

\begin{table}[htb]
\caption{Sample characteristics and parameters of activations at keV neutron energies.
\label{samples}}
\begin{ruledtabular}
\begin{tabular} {lcccc}
 Sample$^a$ 	& Thickness		& Mass 			& Activation	& Average n-flux$^b$	\\
 		 		& (mm)		& (mg) 			& time (h)	&  ($10^{9}$ s$^{-1}$cm$^{-2}$)	\\
\hline
KIT-1M			& 1.8   		& 334.8 			& 251.5 		& 0.84			\\
KIT-2A			& 2.3			& 369.5			& 369.5		& 0.95			\\
Gold foils			& 0.03		& 16.1-16.8		& 			&				\\
				&			& 				&			&				\\
KIT-3M			& 1.4			& 258.5 			& 54.5		& 6.99			\\
KIT-4A			& 1.2			& 188.4			& 44.3		& 5.37			\\
Gold foils			& 0.02		& 11.3-11.8		& 			&				\\
\end{tabular}
\end{ruledtabular}
$^a$ Fe samples pressed from metal powder, gold foils cut from metal sheets; all 
samples 6 mm in diameter.\\
$^b$ Averaged over activation time.
\end{table}

For probing the neutron energy ranges of relevance in AGB stars and in 
massive stars, proton energies of 1912 and 2284 keV were chosen,
respectively. With a proton energy of 1912~keV, 31~keV above the 
threshold of the $^7$Li($p, n$) reaction and using Li layers 30 $\mu$m 
in thickness, kinematically collimated neutrons 
are produced, which are emitted into a forward cone of 120$^\circ$ 
opening angle. Integration over this neutron field yields a quasi-stellar 
Maxwell-Boltzmann (q-MB) spectrum for a thermal energy of 
$kT=25\pm0.$5~keV \cite{RaK88}.

\begin{figure}[hbt]
 \includegraphics[width=7.5cm]{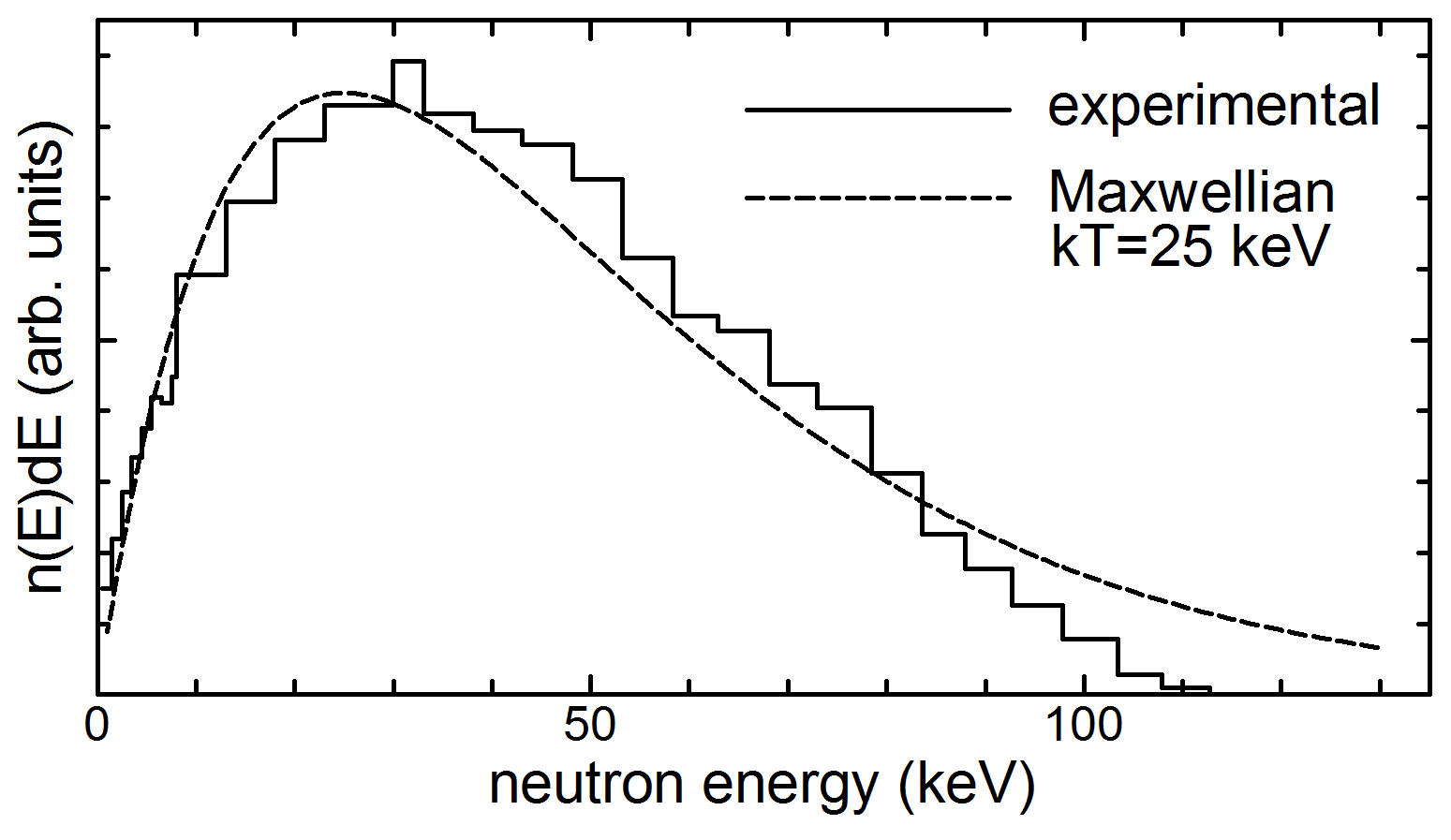}
 \includegraphics[width=7.5cm]{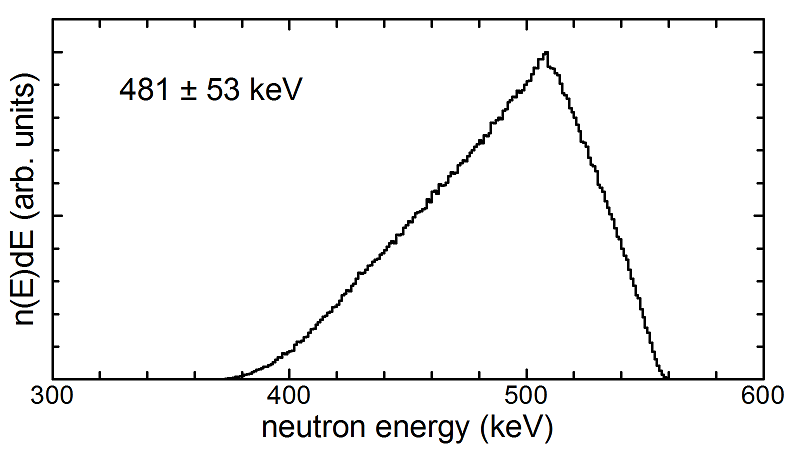}
 \caption{Neutron energy distributions in the irradiations at the Karlsruhe Van de Graaff accelerator obtained  with protons of 1912 keV energy (top) and with 2284 keV (bottom). 
\label{fig:nspectra}}
\end{figure}

Two activations have been carried out for each of the neutron energies.
The main parameters of the irradiations are summarized in Table \ref{tab:irr}.
At the lower energy around 25 keV the sample sandwich was in direct 
contact with the target backing, because the maximal emission angle of 120 $^\circ$ ensured that it was fully exposed to the quasi-stellar field 
(see, e.g. \cite{MDD14,MDD10}) independent of the sample thickness. 
At $E_p=2284$ keV, however, where neutron emission is nearly isotropic, 
a distance of 4 mm was chosen between Li target and sample for restricting 
the energy range of the neutron flux hitting the sample. At this higher proton 
energy 5-$\mu$m-thick Li layers have been used. The resulting neutron 
spectrum centered at $481\pm53$ keV FWHM was calculated with the 
interactive Monte Carlo code PINO \cite{RHK09} with the actual irradiation 
parameters as input. The corresponding neutron spectra are 
plotted in Fig. \ref{fig:nspectra}.

For the gold reference cross section in the energy range of the 25 keV q-MB 
spectrum the prescription of the new version KADoNiS v1.0 \cite{DSP14}
has been followed by adopting the weighted average of recent data from
measurements at GELINA \cite{MBD14} and n$\_$TOF \cite{MDC10,LCD11}. 
This choice is also in perfect agreement with a recent activation measurement 
\cite{JiP14}. Note, the effective values for the 25 keV q-MB spectrum listed in column three of Table 
\ref{tab:irr} are reflecting a change of 5.3\% in the gold reference cross section 
compared to the values previously used in similar activation experiments. 

For the $481\pm53$ keV spectrum, where the ($n, \gamma$) cross section 
of gold is an established standard \cite{CPH11,CPC14}, the evaluated data from 
the ENDF/B-VII.1 library have been used. The respective spectrum averaged 
gold cross sections are listed in Table \ref{tab:irr}. 

\begin{table}[bt]
\caption{\label{tab:irr} Parameters for the irradiations at keV neutron energies.}
\begin{ruledtabular}
\begin{tabular}{cccc}
Sample	& $E_p$			& Gold cross section$^a$	& Neutron fluence 		\\
		&  (keV)			& 	 (mb)				& ($10^{15}$ cm$^{-2}$)\\
\hline \\
\multicolumn{4}{c}{q-MB$^b$ at $kT=25$ keV}							\\
KIT-1M	& 1912 			&	632$\pm$9			& $0.756\pm0.023$		\\
KIT-2A	& 1912	 		& 632$\pm$9			& $1.270\pm0.038$		\\
		& 				&  						& 						\\
\multicolumn{4}{c}{$\overline{E}_n=481\pm53$ keV}						\\
KIT-3M	& 2284 			& $139.0\pm$1.4			& $1.370\pm0.041$		\\
KIT-4A	& 2284			& $139.0\pm$1.4			& $0.857\pm0.026$		\\
\end{tabular}
\end{ruledtabular}
$^a$Spectrum averaged values.\\	
$^b$quasi-Maxwell-Boltzmann distribution simulating a thermal spectrum at $kT=25$ keV.
\end{table}

 \section{AMS measurements \label{sec:4}}

\begin{figure*}[ht]
\includegraphics[width=17cm]{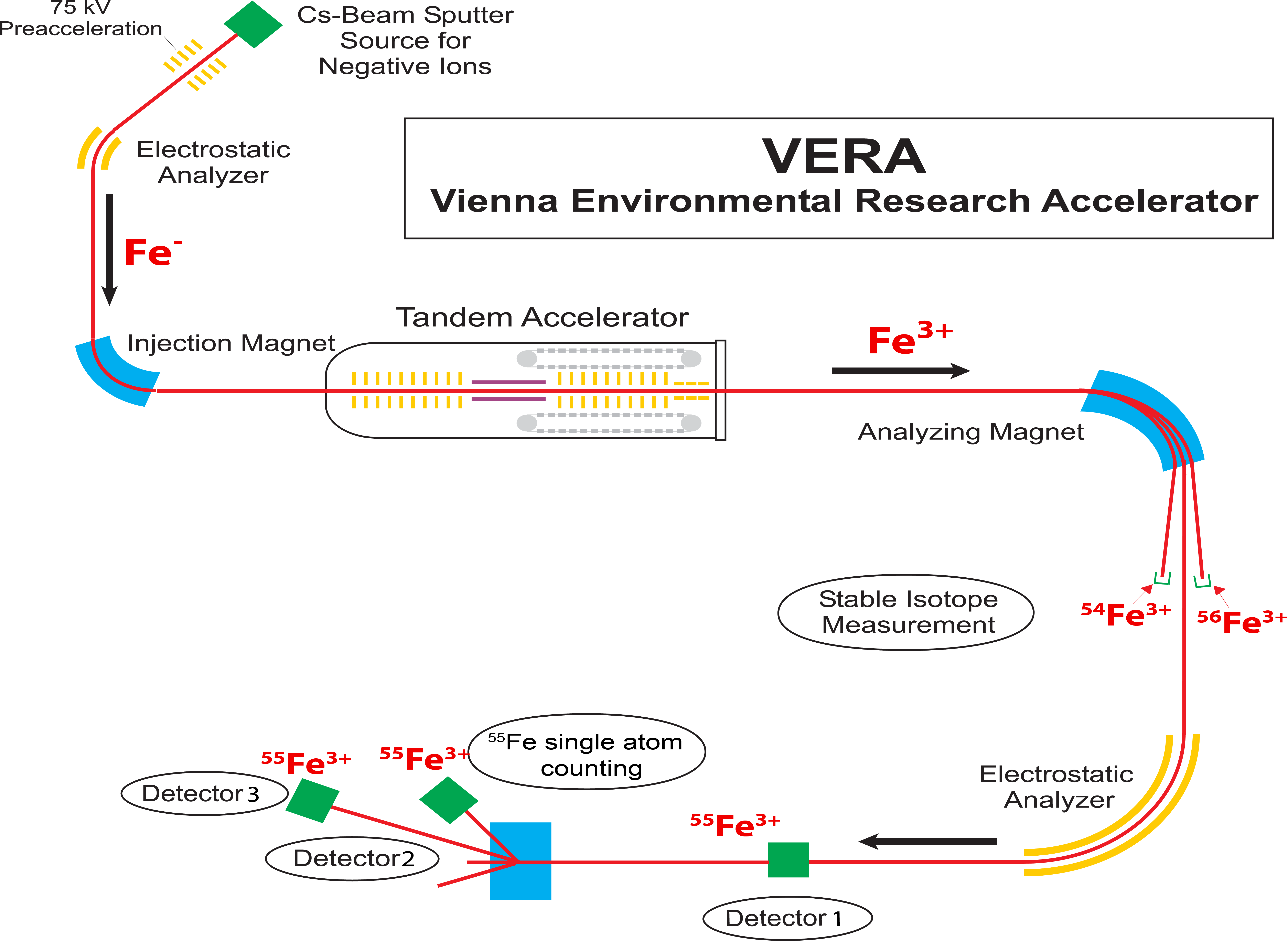}
\caption{Color online. Schematic layout of the AMS facility VERA. Negative Fe ions 
were extracted from the ion source and mass analyzed before the tandem accelerator.
After stripping in the terminal the 3-fold positivley charged (3$^+$) ions  with an energy of 12 MeV were selected for
analysis. The stable $^{54,56}$Fe nuclei were measured with Faraday cups, and the 
rare nuclide $^{55}$Fe was counted in one of the three subsequent particle detectors (see text for details).  
 \label{fig:vera}}
\end{figure*}

As $^{55}$Fe decays almost completely into the ground-state of $^{55}$Mn 
($t_{1/2}$= 2.744(9)~yr, with only $1.3\times10^{-7}$ $\gamma$-rays per decay), the 
$^{55}$Fe nuclei were directly counted \-- prior to their decay to stable $^{55}$Mn\-- by AMS measurements at the Vienna 
Environmental Research Accelerator (VERA), a state-of-the-art AMS facility based 
on a 3-MV tandem \cite{SGL05, WBB14}. A schematic view of the VERA facility 
is shown in Fig. \ref{fig:vera} including the detection devices for recording the 
stable $^{54,56}$Fe and the low-intensity $^{55}$Fe ions.

Negatively charged Fe ions from a cesium sputter source are pre-accelerated and mass-analyzed 
in a low energy spectrometer. In the extracted beam isobaric background due to $^{55}$Mn 
was completely suppressed, because $^{55}$Mn does not form stable negative ions \cite{KMF90}. For 
Fe ions the terminal voltage of the tandem accelerator was set to 3 MV. 
Remaining molecular beam impurities are completely destroyed in the terminal stripper, 
thus eliminating any isobaric interferences with the subsequent mass-selective filters  (see Fig. \ref{fig:vera}). After acceleration ions with charge 3$^+$ 
and an energy of ~12 MeV were selected in the analyzing magnet. The stable $^{54,56}$Fe ions 
were counted as particle currents with Faraday cups, whereas the low intensity $^{55}$Fe 
fraction in the beam was subjected to further background suppression by the electrostatic 
analyzer and was eventually recorded with one of the energy detectors. 

The isotopes $^{56}$Fe, $^{54}$Fe, and $^{55}$Fe were sequentially injected as negative ions into the accelerator. By rapidly varying 
the respective particle energies the different masses of the Fe isotopes were accomodated resulting in the same mass-energy product, this is the particles were adjusted to the same magnetic rigidity at the injection magnet (so-called beam sequencer, not shown in Fig. \ref{fig:vera}).
The stable Fe isotopes were analyzed by current 
measurements with Faraday cups after the injection magnet and after the analyzing magnet 
(for $^{56}$Fe and $^{54}$Fe, respectively). The beam intensity of $^{55}$Fe was measured as countrate with one of the particle detectors. 
This sequence was repeated 5 times per second with millisecond injection 
times for $^{54,56}$Fe, whereas the remaining 95\% of the time were used for $^{55}$Fe 
counting. The transmission through the accelerator was monitored by the currents measured 
at the low- and the high energy side. 
Because the measured $^{54}$Fe and $^{56}$Fe currents are defined by the isotopic 
composition of natural iron, the AMS runs of standards and irradiated samples could be based on both, the $^{54}$Fe and 
the $^{56}$Fe beam. 

The $^{55}$Fe/$^{56}$Fe ratio produced in the irradiations of
typically 10$^{-12}$ was recorded with a background of less than $2\times10^{-15}$. 
Accordingly, the background contributes only less than 0.3 counts per hour to the observed 
$^{55}$Fe count rate of about one every few seconds. 
Under these conditions, a reproducibility of 1\% could 
be reached \cite{WBD07,Wal10, WBB13}. 

\begin{figure}[hbt]
 \includegraphics[width=10cm]{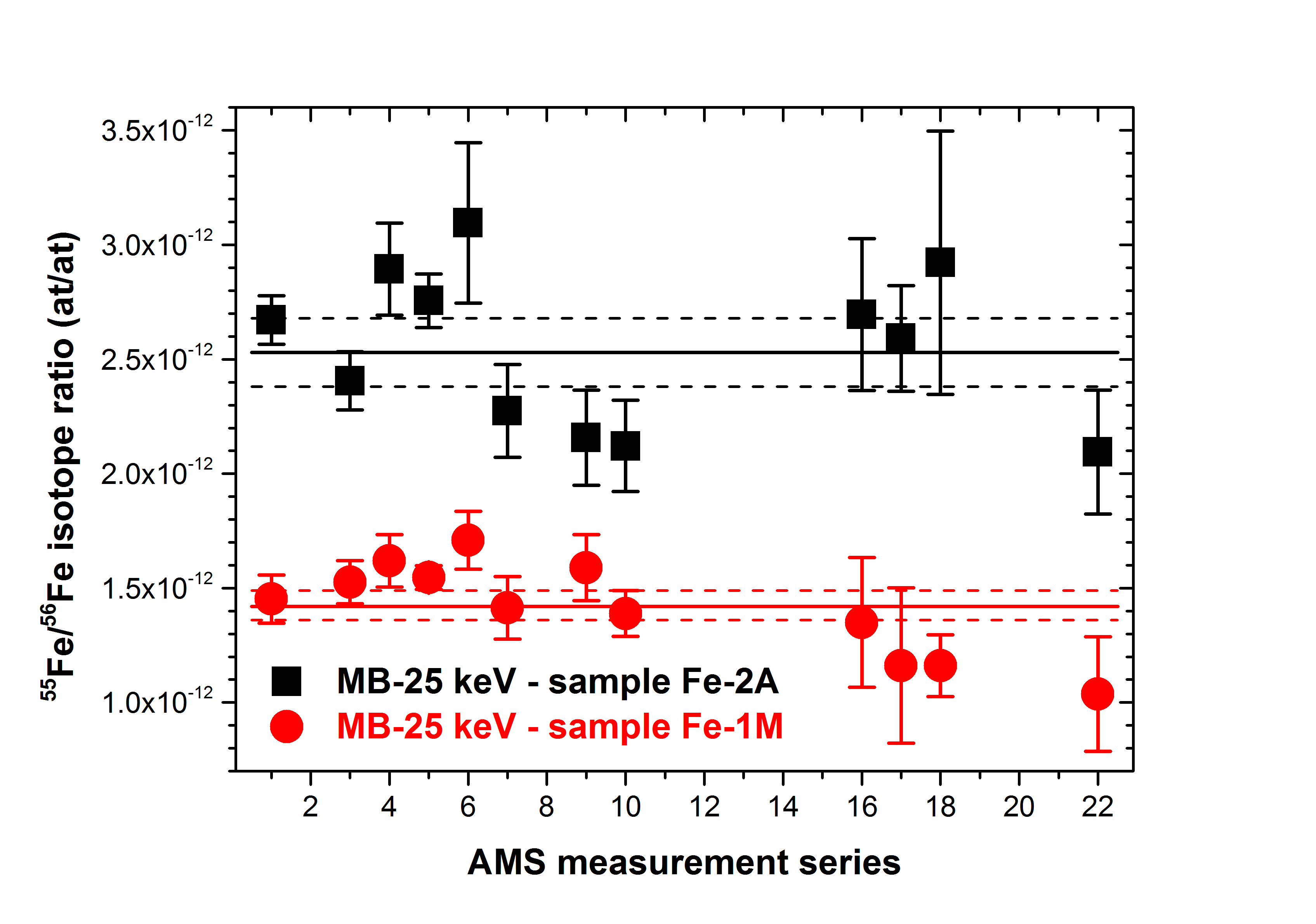}
 \includegraphics[width=10cm]{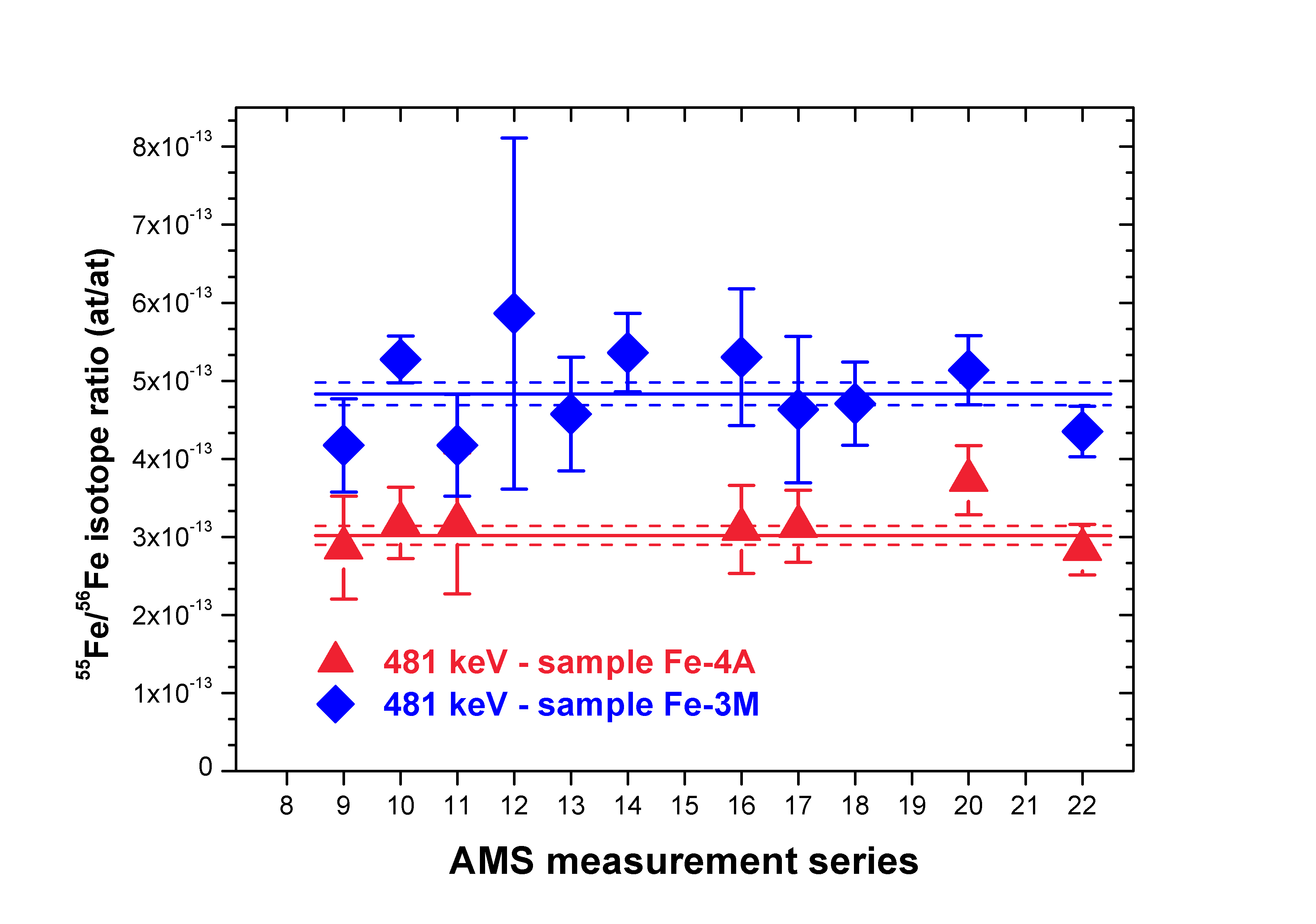}
 \caption{Mean $^{55}$Fe/$^{56}$Fe ratios as obtained during the various AMS beam times. The upper panel gives the data for the two samples irradiated at KIT with a quasi-stellar Maxwell-Boltzmann spectrum, the lower panel represents the data for the two samples irradiated at KIT with higher energies around 481 keV. The solid and dashed lines represent the weighted mean and the standard deviation of the mean for the respective samples.  
\label{fig:AMS_isotope_ratios}}
\end{figure}

The $^{55}$Fe/$^{56}$Fe ratios from the irradiations at KIT as measured during the various AMS beam times are plotted in Fig. \ref{fig:AMS_isotope_ratios}. The upper panel gives the data for the two samples activated in the quasi-stellar Maxwell-Boltzmann spectrum, the lower panel represents the data for the two samples activated at the higher energies around 481 keV. The solid and dashed lines represent the weighted mean and the standard deviation of the mean for the respective samples. All data are corrected for decay of $^{55}$Fe since their production in the various activations.

Because inherent effects such as mass fractionation, machine instabilities, or potential beam 
losses between the current measurement and the respective particle detector are difficult to 
quantify in an absolute way to better than 5 to 10\%, accurate AMS measurements depend 
on well-defined reference materials. Therefore, the isotope ratios $^{55}$Fe/$^{54,56}$Fe 
have been measured relative to an $^{55}$Fe/$^{54,56}$Fe standard produced by means of an $^{55}$Fe 
reference solution by the German metrology laboratory at PTB Braunschweig, with a certified 
1-$\sigma$ uncertainty of $\pm1.5$\% \cite{B09,WBB13}. Details on the AMS procedure for $^{55}$Fe measurements are 
given in Refs. \cite{WBD07,Wal10, WBB13}.

\section{Data analysis and results \label{sec:5}}

\subsection{Neutron fluence \label{sec:5.1}}

The induced activities of the Au (activations at KIT and BNC) and Zr (activation at ATI) monitor foils were measured 
using high-purity germanium (HPGe) detectors. The $\gamma$ efficiency was calibrated 
with a set of accurate reference sources and was known with an uncertainty of 
$\pm$2.0\%. The corrections due to coincidence summing and sample extension were 
minimized by keeping the distance between sample and detector much larger than the 
respective diameters. 

The number of counts $C$ in the characteristic 411.8 keV line in the Au
$\gamma$-ray spectrum recorded during the measuring time $t_{m}$ 
is related to the number of produced nuclei $N_{198}$ at the end of irradiation by 
\begin{equation}
\label{counts}
N_{198}=\frac{C}{K_{\gamma}\epsilon_{\gamma}I_{\gamma}(1-e^{-\lambda t_{m}})e^{-\lambda t_{w}}}
\end{equation}
\noindent
where $\epsilon_{\gamma}$ denotes the detector efficiency and $t_{w}$ the waiting 
time between irradiation and activity measurement. The decay rate $\lambda=0.25728(2)$ 
d$^{-1}$ and the intensity per decay, $I_{\gamma}=95.62(6)$\% of $^{198}$Au were 
adopted from Ref. \cite{XiM16}. The factor $K_{\gamma}$ describes the $\gamma$-ray self 
absorption in the sample, which is for the thin gold samples used in this work in very good 
approximation \cite{Dix51}
\begin{equation}
\label{self-absorption}
K_{\gamma}=\frac{1-e^{-\mu x}}{\mu x}.
\end{equation}
\noindent
The $\gamma$-ray absorption coefficient $\mu$ was taken from Ref.~\cite{HuS09}. 
The gold samples were $x=20-30$ $\mu$m in thickness, yielding self-absorption 
corrections $K_{\gamma}=0.996-0.994$. Accordingly, the associated uncertainties had 
no effect on the overall uncertainty budget and were neglected.  

Similarly, the number of produced $^{95}$Zr nuclei in the ATI irradiations was deduced from the decay rate 
 $\lambda=0.0108(1)$ d$^{-1}$ and the intensity per decay, $I_{\gamma}=44.27(22)$\% for 724.2 keV and $I_{\gamma}=54.38(22)$\% for 756.7 keV 
(adopted from Ref. \cite{BMS10}).

The number of produced nuclei $N_{198}$ or $N_{95}$ ($N_{prod}$) can also be expressed by the neutron fluence 
$\Phi_{tot} = \int^{t_{a}}_{0} \Phi (t)dt$, the corresponding spectrum-averaged 
capture cross section $\langle \sigma \rangle$, the decay correction $f_b$, and the 
number of irradiated atoms in the sample $N$ as
\begin{equation}
\label{activated-nuclei}
N_{prod}=\Phi_{tot} N \langle \sigma \rangle f_{b}.
\end{equation}
\noindent

The factor $f_{b}$, which corrects for the fraction of activated nuclei that decay 
already during irradiation, is 
\begin{equation}
\label{fb-factor}
f_{b}=\frac{\int^{t_{a}}_{0}\Phi (t)e^{-\lambda(t_{a}-t)}dt}{\int^{t_{a}}_{0} \Phi (t)dt} 
\end{equation}
\noindent
where $\Phi (t)$ denotes the neutron intensity during the irradiation and $\lambda$ 
the decay rate of the product nucleus $^{198}$Au or $^{95}$Zr.

In the short activations at ATI this correction is almost negligible 
because the half-lives of the activation product $^{95}$Zr was much longer than the 
irradiation times $t_{a}$. In the longer irradiations at BNC and KIT it had to be considered 
for the gold activities, where the half-life of $t_{1/2}=2.6941(2)$ d is shorter than 
the irradiation times of about 1 and 4 days. Due to the constant neutron flux provided by 
the reactor, $f_b$ can be determined by integrating Eq. \ref{fb-factor}.

In the ATI activations, the accuracy of the fluence was limited 
by the epithermal correction for the $^{94}$Zr monitors. The total production of $^{95}$Zr consists of the thermal part (49.4 mbarn) and the epithermal part (280 mbarn) with the epithermal flux only 1/76 of the thermal flux for this irradiation setup (see above). The measured total $^{95}$Zr activitiy was corrected for the additional $7\pm4$ \% epithermal production and from that the thermal neutron fluence was calculated (Tables \ref{tab:irrth} and \ref{tab:irrth_res}). Since the ratio of the epithermal to the thermal cross section is much lower for the $^{54}$Fe case, the equivalent correction for  $^{54}$Fe($n, \gamma$)$^{55}$Fe was 1\%. 
In the end the fluence for the ATI samples could be determined with an uncertainty of about 5\%.

For the cold neutron beam at the BNC the neutron spectrum is characterized
 by a pure 1/v-shape with energies below 50 meV \cite{BKS14}. As also the cross sections 
of $^{197}$Au and $^{54}$Fe exhibit a 1/v-shape in this energy range, 
the reaction rates are scaling in exactly the same way from cold to thermal 
energies. Accordingly, there are no corrections for epi-thermal neutrons in 
this case. In addition, these irradiations were performed in a well-defined 
geometry with the sample stack mounted perpendicular to the neutron beam. 
By comparison of the activities of the gold powder mixed with Fe in the pellets with the front and back foils in the stack it could be 
demonstrated that the respective fluence values were consistent within 
1\%, thus constraining possible corrections for inhomogeneities of the beam 
and scattering effects. The effective fluence could be derived 
with an accuracy of 2\% as detailed in Table \ref{tab:error_th}.

\begin{table}[bt]
\caption{\label{tab:error_th} Uncertainty contributions for the thermal cross 
section value.}
\begin{ruledtabular}
\begin{tabular}{lc}
Source of Uncertainty 							& Uncertainty (\%)  				\\
\vspace {-3mm} \\
\hline
\underline{Neutron fluence} 						&	ATI / BNC$^a$							\\
\hspace*{5mm} - Zr / Gold cross section		& 3.4 / 0.1 							\\
\hspace*{5mm} - Epithermal correction			& 4 / 0$^a$						\\
\hspace*{5mm} - Mass of Zr / Gold samples		& $<0.2$ / $<0.1$						\\
\hspace*{5mm} - $\gamma$ efficiency 			& 2.0							\\
\hspace*{5mm} - $\gamma$ intensity per decay 	& 0.5 / 0.12	 						\\
\vspace {1mm}
\hspace*{5mm} - Time factors				& $<0.1$ 						\\			
\hline       														\\			
\vspace {-5mm}  \\
\hspace*{5mm} - total fluence:				& 5.6 / 2.0 \% 						\\			
\hline       														\\	
\underline{AMS measurement}						&								\\
\hspace*{5mm} - PTB standard				& 1.5							\\	
\hspace*{5mm} - Atom counting 				& $<$1							\\	
\hspace*{5mm} - $^{56}$Fe current			& 0.6							\\			
\hspace*{5mm} - AMS reproducibility			& 1.5							\\			
\hline       														\\			
\vspace {-5mm}  \\
\hspace*{5mm} - total AMS:				& 2.5 \% 						\\			
\hline       														\\
Total:	ATI (thermal) / BNC (cold)				& 6.4 / 3.2
\end{tabular}
\end{ruledtabular}
$^a$ For thermal and cold neutrons, respectively.
\end{table}

For the Karlsruhe activations at keV energies the effective gold reference cross 
section had to be determined by folding with the experimental neutron energy 
distributions, i.e. the quasi-MB spectrum at $kT=25$ keV and the spectrum
around $481\pm53$ keV. The cross-section of the $^{197}$Au($n, \gamma$) 
reaction was adopted according to the recommendation in the new version KADoNiS 
v1.0 \cite{DSP14} and yields 
spectrum-averaged Au cross sections of $^{197}$Au  with uncertainties of 
1.5 and 1\% (Table \ref{tab:irr}).  

The keV-neutron flux produced at the Karlsruhe Van de Graaff showed considerable 
non-uniformities due to the decreasing performance of the Li targets as well as to
fluctuations in the beam intensity. Therefore, the correction factor $f_{b}$ had to 
be evaluated by numerical integration of Eq.~\ref{activated-nuclei} using the 
time-dependence of the neutron flux that was recorded by the $^6$Li-glass detector
as mentioned above. 

The main contributions to the total 3\% uncertainty of the neutron fluence are 
due to the $\gamma$ efficiency of the HPGe detector and to the Au reference 
cross sections (Table \ref{tab:uncertainties}).
\begin{table}[bt]
\caption{\label{tab:uncertainties} Uncertainty contribution for the spectrum averaged cross 
sections at keV energies.}
\begin{ruledtabular}
\begin{tabular}{lc}
Source of Uncertainty 							& Uncertainty (\%)  				\\
\hline
Neutron fluence 								&								\\
\hspace*{5mm} - Gold cross section			& 1.5/ 1.0 $^a$					\\
\hspace*{5mm} - Mass of gold samples			& 0.3							\\
\hspace*{5mm} - $\gamma$ efficiency 			& 2.0							\\
\hspace*{5mm} - $\gamma$ intensity per decay 	& 0.12	 						\\
\hspace*{5mm} - Time factors					& $<0.1$ 						\\			
											&								\\
AMS measurement							&								\\
\hspace*{5mm} - PTB standard				& 1.5							\\	
\hspace*{5mm} - Atom counting 				& $<$2							\\	
\hspace*{5mm} - $^{56}$Fe current			& 0.6							\\			
\hspace*{5mm} - AMS reproducibility			& 1.5							\\			
											&								\\
Total										& 4.0/ 3.8
\end{tabular}
\end{ruledtabular}
$^a$ For neutron energies of 25 (q-MB) and 481 keV, respectively.
\end{table}

\subsection{Spectrum-averaged cross sections \label{sec:5.2}}

The  spectrum-averaged $^{54}$Fe($n, \gamma$) cross section can directly be 
calculated from the total neutron fluence $\Phi_{\rm tot}$, and the isotope ratio 
$^{55}$Fe/$^{56}$Fe and $^{55}$Fe/$^{54}$Fe measured via AMS, 
\begin{equation}
\langle \sigma \rangle = \frac{1}{\Phi_{\rm tot}}\times \frac{^{55}{\rm Fe}}{^{54}{\rm Fe}} = \frac{1}{\Phi_{\rm tot}}\times \frac{^{55}{\rm Fe}}{^{56}{\rm Fe}}\times\frac{^{56}{\rm Fe}}{^{54}{\rm Fe}}
\end{equation}
where  $^{56}$Fe/$^{54}$Fe= $N_{56}$/$N_{54}$=$15.70\pm0.09$ corresponds to the natural isotope ratio 
in the sample (91.75\% and 5.85\%, respectively) \cite{BeW11}. Note the particular advantage of the AMS method, i.e. that 
the cross section is determined completely independent of the sample mass and the decay 
properties of the product nucleus. 

\subsubsection{Thermal cross section \label{sec:5.2.1}}

\begin{table}[htb]
\caption{Results from the activations at cold and thermal neutron energies.
\label{tab:irrth_res}}
\begin{ruledtabular}
\begin{tabular} {lccc}
Sample	& 	Neutron fluence $^a$ 		&  $^{55}$Fe/$^{54}$Fe   	& Thermal cross  		\\
		& (10$^{13}$ cm$^{-2}$)		&	(10$^{-10}$ at/at)		& section (b) 		\\
\hline
ATI-FeM	& $21.2\pm1.1$				& $4.69\pm0.03$			& $2.21\pm0.15$		\\
ATI-Fe2	& $22.4\pm1.2$				& $5.10\pm0.07$			& $2.28\pm0.15$		\\
ATI-FeA2	& $4.59\pm0.35$				& $1.03\pm0.022$			& $2.24\pm0.15$		\\
ATI-FeA4	& $2.65\pm0.14$				& $0.61\pm0.02$			& $2.30\pm0.15$		\\
mean		&						&					& $2.26\pm0.15$		\\				
		&						&					& 				\\				
BNC-FeM	& $1.042\pm0.025$			& $0.246\pm0.008$		& $2.36\pm0.07$		\\
BNC-Fe4	& $0.396\pm0.009$			& $0.091\pm0.002$		& $2.29\pm0.06$		\\
mean		&						&					& $2.31\pm0.07$		\\
		&						& 												\\				
\multicolumn{3}{l}{Weighted mean}								& $2.30\pm0.07$ 	\\
\end{tabular}
\end{ruledtabular}
$^a$ Thermal equivalent neutron fluence (ATI values include 7\% correction for epithermal contribution
of Zr monitor foils).
\end{table}

\begin{center}
\begin{figure}[hbtp]
\hspace*{-1.2cm}
\includegraphics[width=11cm]{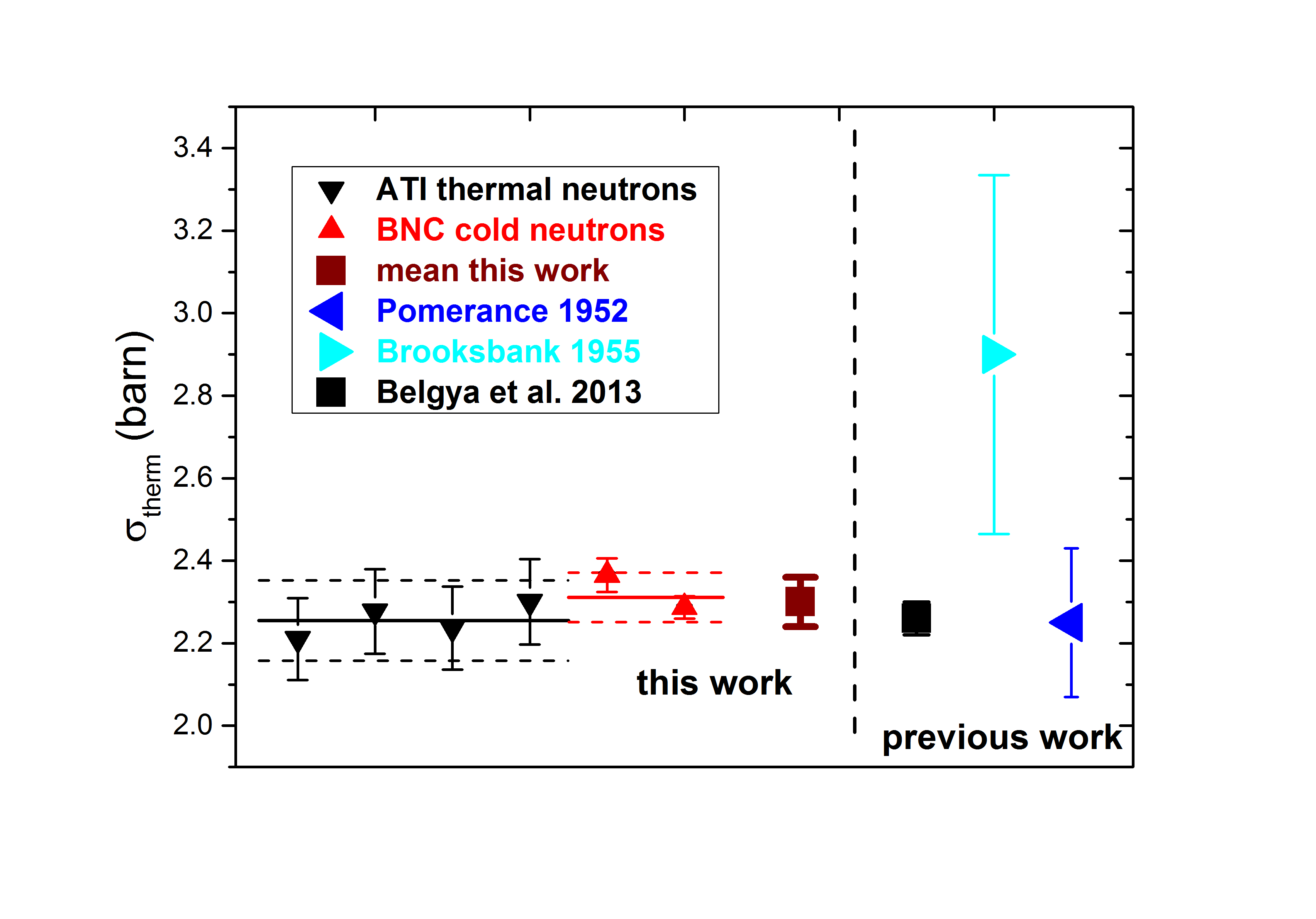}
\caption{(Color online) Comparison of the thermal cross section values obtained in this work from two independent activations with thermal (ATI) and cold (BNC) neutrons. Also plotted are the two previous experiments from more than 60 years ago by Brooksbank {\it et al.} \cite{BLL55} and by Pomerance \cite{Pom52}. The weighted average (square) of our work is in very good agreement with the value of \cite{Pom52} that was the basis for the recommend value in \cite{Mug06}. Also plotted is a new value quoted by Belgya et al. that is based on an improved decay scheme of $^{55}$Fe\cite{BKS13}.
\label{fig:thermal exp data comp previous}}
\end{figure}
\end{center}

Both thermal cross section values deduced from the BNC and ATI activations \cite{B09} show a very good agreement (see Fig. \ref{fig:thermal exp data comp previous}). The scatter of the individual results is small ($\pm$2\%) and the final uncertainty is dominated by systematic contributions. The weighted mean gives our thermal cross section value of $2.30\pm0.07$ b, which is well compatible
with the value of $2.25\pm0.18$ b recommended in Ref. \cite{Mug06}, but a factor of 2.5 more accurate (see Table \ref{tab:irrth_res}). Our data fit also well to the recently published value of Belgya et al. that is based on an improved knowledge of the decay scheme of $^{55}$Fe, resulting in a thermal cross section of $2.29\pm0.05$ b \cite{BKS13}.

\subsubsection{Cross sections at keV energies \label{sec:5.2.2}}

The measured $^{55}$Fe/$^{56}$Fe ratios are listed in Table \ref{tab:results} together with
the resulting spectrum averaged cross sections. The uncertainties associated with the AMS 
measurement are determined by the $^{55}$Fe standard (1.5\%), the $^{56}$Fe current 
(0.6\%), the reproducibility of the AMS runs (1.5\%), and the counting of the unstable 
$^{55}$Fe nuclei (3 - 6\% for individual AMS runs). The statistical uncertainties become 
$<$2\% when all AMS beam times are combined. The quadratic sum of these contributions 
yields an effective AMS uncertainty of 3\% (Table \ref{tab:uncertainties}).

\begin{table}[htb]
\caption{Results for the activations at keV neutron energies.
\label{tab:results}}
\begin{ruledtabular}
\begin{tabular} {ccccc}
Energy		& Sample 	& $^{55}$Fe/$^{56}$Fe 	 	& $\langle$$\sigma$$\rangle$$^a$	& ENDF/B$^a$ 	\\
(keV)		&			&	($10^{-12}$)			& (mb)				&	-VII.1	\\
25 (qMB) 	& KIT-1M	& 	$1.42\pm0.05$			& $31.3\pm1.5$		& 		\\
 			& KIT-2A	&	$2.53\pm0.11$			& $29.6\pm1.4$		& 		\\
			& Adopted	&							& $30.3\pm1.2$		& 22.6 	\\
			&			&							&					& 		\\
$481\pm53$	& KIT-3M  	& 	$0.48\pm0.01$			& $6.01\pm0.28$		& 		\\
			& KIT-4A 	&	$0.30\pm0.01$			& $6.02\pm0.23$		& 		\\
			& Adopted	&							& $6.0\pm0.2$		& 7.3	\\
\end{tabular}
\end{ruledtabular}
$^a$ Spectrum averaged cross sections.
\end{table}

The comparison in Table \ref{tab:results} shows that the present results are 25\% 
higher at 25 keV and 20\% lower at 481 keV than obtained by folding the evaluated 
cross section from the ENDF/B-VII.1 library with the respective neutron spectra. 
While the evaluated data imply a rather weak energy dependence, the present 
results are consistent with a 1/$\sqrt{E_n}$ dependence on energy, in full 
agreement with the cross section shape implied by the experimental TOF data 
\cite{Giu14,GDT14,BCR83,AMB77,AMB79}. Therefore, this energy trend is to be 
preferred for improving the MACS values (see Sec. \ref{sec:astro1}).

For comparison with the present result, the TOF data were averaged over the 
25 keV-qMB distribution $N(E)$ using the approximation \cite{MaG65}
 \begin{equation}\label{eq:macklin}
\langle \sigma \rangle =  \frac{\int \sigma_{th} \sqrt{\frac{25\times10^{-6}}{E}} N(E) dE+ 
\sum_i N(E)_i A_{\gamma, i}}{\int N(E) dE},
\end{equation}
where the first term in the nominator represents the 1/$\sqrt{E_n}$ extrapolation of our new 
thermal cross section value $\sigma_{th} = 2.30\pm0.07$ mb (see Sec. \ref{sec:5.2}). The resonance 
contribution is obtained by the sum of the resonance areas 
$$A_{\gamma,i}=\frac{2\pi^2}{k_n^2} \frac{g\Gamma_\gamma\Gamma_n}{\Gamma_\gamma+\Gamma_n}$$ 
which are determined by the radiative and neutron widths $\Gamma_\gamma$, 
$\Gamma_n$,  the wave number $k_n=2.1968\times10^9 \times A/(A+1) \sqrt{E_n}$, 
and the statistical factor  $g=(2J+1)/(2I+1)(2s+1)$. With this approximate prescription, 
the resonance parameters of Giubrone \cite{Giu14}, Brusegan {\it et al.} \cite{BCR83}, 
and Allen {\it et al.} \cite{AMB77,AMB79} yield spectrum-averaged cross sections of 
30.9, 30.5, and 32.8 mb, respectively. The weighted average of 31.3$\pm$2.1 mb is 
about 3\% higher than our value of 30.3$\pm$1.2 mb, well within uncertainties. 

An additional test was made using the resonance parameters of Giubrone \cite{Giu14}. The contributions of the broad s-wave resonances have been expressed by a sum of 
Breit-Wigner terms, 
$$ \sigma(E_n) = \frac{\pi}{k_n^2} \frac{g\Gamma_\gamma \Gamma_n}{(E_n -E_{res})^2 
+ (\Gamma_\gamma + \Gamma_n)^2/4},$$
yielding a partial spectrum average of 9.6 mb. As the low-energy tails 
of these resonances contribute already a fraction of 834 mb to the thermal cross section, the 
$1/v$-extrapolation from thermal to 25 keV is reduced from 2.08 to 1.24 mb. The narrow 
resonances with $\ell$$>$0 can again be treated as a weighted sum of the resonance areas and
are found to contribute another 19.3 mb. In total, the Breit-Wigner approach gives 30.1 mb, 
in fair agreement with the 30.9 mb obtained via Eq. \ref{eq:macklin}, thus justifying the use
of this expression \cite{BVW92}.

At this point it is interesting to note that the refined experiments \cite{Giu14,BCR83} 
yield spectrum averaged cross sections in significantly better agreement with the present 
result than the first attempt described in \cite{AMB77,AMB79}. In fact, within uncertainties 
these values are consistent, indicating the proper treatment of neutron backgrounds in the 
analysis of the broad s-wave resonances, especially those at 7.8, 52.8, and 99.1 keV
neutron energy (see Fig. \ref{fig:macs30_comp}).

\section{Astrophysical aspects \label{sec:astro}}

\subsection{Maxwellian averaged cross sections \label{sec:astro1}}

In view of the difficulties with the energy dependence of the evaluated cross 
section \cite{CHO11}, additional MACS values have been calculated from the 
experimental resonance data of Refs. \cite{Giu14,BCR83,AMB77} using the 
approximation of Macklin and Gibbons \cite{MaG65} 
\begin{eqnarray*}
\label{eq:macklin_org}
\frac{\langle \sigma v \rangle}{v_T} 
= \sigma_{th} \sqrt{\frac{25\times10^{-6}}{kT}} +
\frac{2}{\sqrt\pi}\frac{1}{(kT)^2}\sum_i A_{\gamma, i}E_i\exp{(\frac{-E_i}{kT})}
\end{eqnarray*}
where $E_i$ denotes the resonance 
energy and $kT$ the thermal energy. As the sum in this 
equation ends at the maximum resonance energy of a given data set, the 
thermal spectrum is truncated at this energy. In order to keep the error caused by the 
truncation close to the experimental uncertainties, MACS values derived from the data in 
Refs. \cite{BCR83,AMB77} have been limited to thermal energies below $kT=60$ keV.

\begin{figure}[hbt]
\includegraphics[width=9cm]{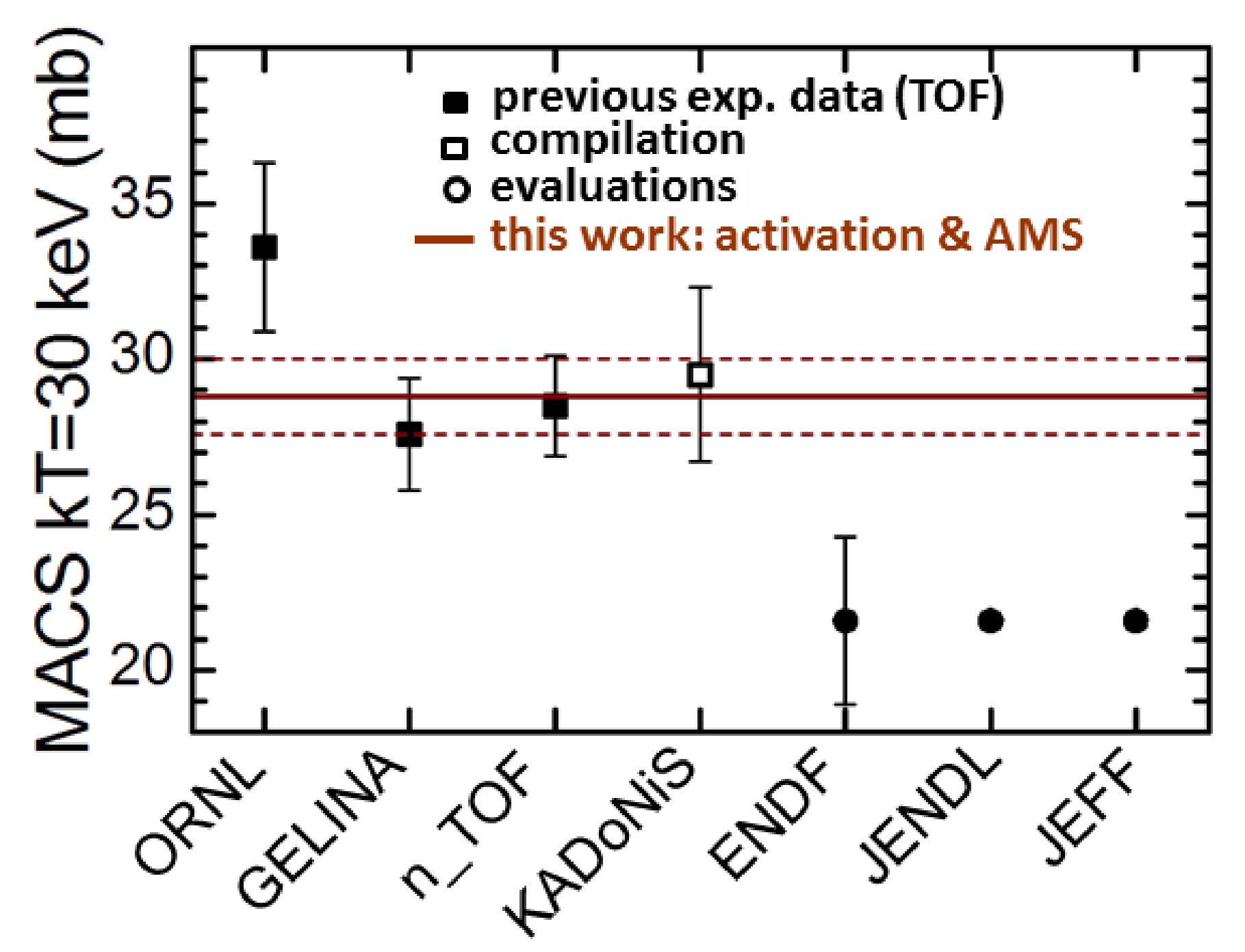}
\caption{Comparison of the present MACS value at $kT=30$ keV with the recommended value in the compilation of Ref. \cite{DPK14}, data obtained in previous TOF measurements (\cite{AMB79,BCR83,Giu14,GDT14}), and calculated from the evaluated cross sections in the ENDF/B-VII.1 \cite{CHO11}, JENDL-4.0 \cite{SIN11}, and JEFF-3.2 \cite{JEF14} libraries.  
\label{fig:macs30_comp}}
\end{figure}

The comparison of the present MACS for $kT=30$ keV in Fig. \ref{fig:macs30_comp} shows good agreement with the refined TOF measurements performed at Geel \cite{BCR83} and at CERN/n\_TOF \cite{Giu14,GDT14}, whereas the evaluated cross sections in the ENDF/B-VII.1 \cite{CHO11}, JENDL-4.0 \cite{SIN11}, and JEFF-3.2 \cite{JEF14} libraries are yielding incompatibly small values. The MACS in the KADoNiS \cite{DPK14} compilation is obviously biased by the high value from Ref. \cite{AMB77,AMB79}.
	
\begin{figure}[ht]
\includegraphics[width=8.cm]{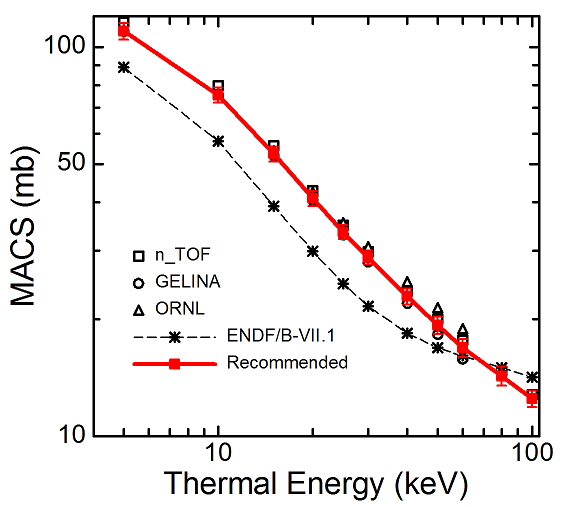}
\caption{(Color online) Recommended MACS values between $kT$=5 and 100 keV compared to data obtained with 
evaluated cross sections from ENDF/B-VII.1 \cite{CHO11} and from TOF-based
experimental data \cite{Giu14,BCR83,AMB77} (see text for details). 
 \label{fig:macs}}
\end{figure}

The temperature dependence of these results (Fig. \ref{fig:macs}) shows that the TOF data are providing a  consistent trend, in accordance with the present results. In contrast, the trend obtained with the evaluated cross sections is clearly overestimating the MACS values above about 30 keV. Therefore, 
the temperature trend defined by the experimental TOF data sets has been adopted for the recommended MACS values in Table \ref{tab:macs}. 

These recommended MACS values are based on the adopted temperature 
trend, but are normalized to the measured spectrum-averaged 
cross section at 25 keV by a factor \\
\centerline{$NF=(30.3\pm1.2 {\rm mb})/31.5 {\rm mb}=0.968\pm0.040$}\\
where the denominator represents the corresponding mean value derived from the TOF 
measurements as described above. The uncertainties are composed of contributions from 
the measured spectrum-averaged cross section ($\pm4.0$\%) and from the energy trend, 
which was estimated via the differences among the TOF-based MACS data (0 to 4.5\%).    

\begin{table}[htb]
\caption{Recommended MACS values from this work (in mb) compared to data obtained with 
evaluated cross sections from ENDF/B-VII.1 \cite{CHO11} and from the KADoNiS v0.3
compilation \cite{DPK09,DPK14}. 
\label{tab:macs}}
\begin{ruledtabular}
\begin{tabular} {cccc}
$kT$ (keV)	& ENDF/B-VII.1	& KADoNiS			& This Work		\\ 
\hline
5			& 88.8 			& $119$		& $102\pm5$	\\
10			& 57.3 			& $81$		& $69.9\pm3.1$	\\
15			& 39.1 			& $56$		& $49.3\pm2.2$	\\
20			& 29.9			& $43$		& $37.9\pm1.7$	\\
25			& 24.7			& $35$		& $31.1\pm1.3$	\\
30			& 21.6			& $29.6\pm1.3$		& $26.7\pm1.1$	\\
40			& 18.4			& $23.9$		& $21.2\pm1.0$	\\
50			& 16.9			& $20.9$		& $17.9\pm0.9$	\\
60			& 16.1			& $19.2$		& $15.7\pm0.8$	\\
80			& 15.0			& $17.4$		& $13.3\pm0.7$	\\
100			& 14.2			& $16.4$		& $11.6\pm0.6$	\\
\end{tabular}
\end{ruledtabular}
\end{table}

\subsection{Nucleosynthesis \label{sec:astro2}}

Among the stable Fe and Ni isotopes, $^{54}$Fe and $^{58}$Ni are unique, 
because they are not produced but depleted via neutron capture, and were, 
therefore, proposed for constraining the neutron exposure of the weak $s$ 
process in the He and C burning zones of massive stars \cite{WoW95}. 

The effect of the new stellar cross section on nucleosynthesis in AGB stars 
was investigated with stellar models of initial mass 2, 3 and 6 $M_\odot$ 
for solar metallicity (Z=0.014) and roughly 1/10$^{\rm th}$ of solar (Z = 
0.001). The effect of the new $^{54}$Fe($n, \gamma$)$^{55}$Fe cross 
section was tested using a 77 species network, which includes a small network around the iron group elements.
For one model (3 $M_\odot$, Z = Z$_{solar}$) a full $s$-process network  that includes species up to Po was used to test the validity of the 77 species runs. Details of the nuclear network and the numerical method employed 
in the post-processing code are given in \cite{KaL16,WBB16}, and 
information on the stellar evolutionary sequences used as input into the 
post-processing can be found in \cite{Kar14,FKL14}. 

During the post-processing we artificially included a proton profile in the 
He-rich intershell at the deepest extent of each dredge-up in the 2 and 3 
$M_\odot$ models. The proton abundance is chosen such that it decreases 
exponentially from the envelope value of $\sim$0.7 to a value of 10$^{-4}$ 
at a location in mass 2$\times$10$^{-3}$ $M_\odot$ below the base of the 
envelope. The protons are captured by the abundant $^{12}$C in the envelope 
to form a region rich in $^{13}$C. In between convective thermal pulses, the 
reaction $^{13}$C($\alpha, n$)$^{16}$O burns radiatively in the intershell 
and releases free neutrons, which are captured by Fe-group isotopes including 
$^{54}$Fe. The by far dominant $^{13}$C neutron source is complemented 
by the $^{22}$Ne($\alpha, n$)$^{25}$Mg reaction that is marginally 
activated by the higher temperatures of $\sim$250 MK during the He shell 
flashes. In the 6 $M_\odot$ model we do not include any protons into the 
He-intershell; instead neutrons are only produced by the $^{22}$Ne($\alpha, 
n$)$^{25}$Mg reaction during convective He-shell burning. The higher 
neutron density - owing to peak temperatures exceeding 300 MK in these -
models (e.g., \cite{FKL14}) allows for the $s$-process reaction flow to bypass 
the branching at $^{59}$Fe and to produce the radioactive $^{60}$Fe.  

Apart from the rate of the  $^{54}$Fe($n, \gamma$)$^{55}$Fe reaction all 
the tests were using the same input for the stellar and nuclear physics to 
compare the effect of the new cross sections presented here with that from 
the KADoNiS database. It turned out that the new cross section does not change 
the average surface composition in the winds of any of the stellar models 
considered. For all Fe isotopes we report changes of $<$1\% for all stellar 
models. Also none of the elements heavier than Fe (e.g., $s$-process elements 
such as Sr or Ba) were affected by changing the cross section of the $^{54}$Fe($n, 
\gamma$)$^{55}$Fe reaction. 

Variations in the order of 1\% are well within the uncertainties of the measured Fe 
abundances in presolar grains. Therefore, the abundances obtained by using 
our new MACS of $^{54}$Fe are consistent with the abundances obtained using the 
previous MACS from the literature.

We mentioned that the depletion of $^{54}$Fe can be used to constrain the neutron 
exposure in stellar model calculations. The $\sim$5\% uncertainty of the MACS obtained 
in this work makes the use of the $^{54}$Fe as a diagnostic more robust, whereas 
uncertainties from other nuclear reactions and from stellar physics assumptions, see 
e.g., Refs. \cite{KWG94,TEM07,PGH10}, are now more relevant. Accordingly, to date there 
seems to be no need for further improvement of the $^{54}$Fe($n, \gamma$) cross 
section in stellar nucleosynthesis applications. 

\section{Summary \label{sec:sum}}

The neutron capture reaction $^{54}$Fe($n, \gamma$)$^{55}$Fe represents 
an excellent candidate for comparing different and independent methods for 
cross section measurements. While time-of-flight based techniques provide 
continuous data over a wide energy range, neutron activation of $^{54}$Fe 
combined with AMS detection of $^{55}$Fe at the VERA laboratory, where 
$^{55}$Fe detection was demonstrated to be precise at a level of 1\%, allows 
one to gain information on cross section values for only a few selected neutron 
energies. In this way, the more complicated TOF technique can be checked 
and normalized with AMS data, in particular in cases of reactions with large
scattering/capture ratios. 

The potential of neutron activation and subsequent AMS analysis for accurate
cross section studies has been demonstrated by the present measurements at 
thermal and keV neutron energies. At thermal, the previously recommended 
value was confirmed, but with a 2.5 times reduced uncertainty. The good agreement with 
the results at 25 keV provides evidence for the proper treatment of strong 
scattering resonances in the analysis of advanced TOF measurements. It was 
also shown that the combination of neutron activation and AMS reached an 
accuracy level that is not only competitive but exceeds that of advanced TOF 
measurements. Accordingly, such data are of key importance for normalization
of previous TOF results.   

The impact of the improved cross sections for neutron capture nucleosynthesis 
was investigated for the case of AGB stars. Indeed, the expected depletion 
effect of $^{54}$Fe was found to be rather weak for constraining the neutron 
fluence in these stars.\\ \\

\centerline{\bf Acknowledgement}

This work was partly funded by the Austrian Science Fund (FWF), project No.s P20434 and I428, and 
by the Australian Research Council, project no. DP140100136. AK thanks Maria Lugaro for the nuclear network used to perform the s-process calculations. MP acknowledges support to NuGrid from NSF grants PHY 02-16783 and PHY 09-22648 (Joint Institute for Nuclear Astrophysics, JINA), grants PHY-1430152 (JINA Center for the Evolution of the Elements) and EU MIRG-CT-2006-046520, support from the "Lend{\"u}let-2014" Programme of the Hungarian Academy of Sciences and from SNF (Switzerland). 

\newcommand{\noopsort}[1]{} \newcommand{\printfirst}[2]{#1}
  \newcommand{\singleletter}[1]{#1} \newcommand{\swithchargs}[2]{#2#1}

\end{document}